\documentclass{PoS}
\usepackage{amsmath}

\title{SU(3) Flavour Symmetries and CP Violation}

\ShortTitle{SU(3) Flavour Symmetries and CP Violation}

\author{L.~Calibbi\\
        SISSA/ISAS and INFN, I-34013, Trieste, Italy.\\
        E-mail: \email{calibbi@sissa.it}}
\author{\speaker{J.~Jones~P\'erez}\thanks{Supported by the Spanish MICINN and FEDER (EC) Grant No.~FPA2008-02878.}\\
        Departament de F\'{\i}sica Te\`orica and IFIC, Universitat de Val\`encia-CSIC, E-46100, Burjassot, Spain.\\
        E-mail: \email{joel.jones@uv.es}}
\author{A.~Masiero\\
        INFN, Sezione di Padova, via F Marzolo 8, I--35131, Padova, Italy.\\
	Univ. of Padova, Physics Dept. "G. Galieli", Padova, Italy.\\
        E-mail: \email{Antonio.Masiero@pd.infn.it}}
\author{J.-h.~Park\\
        INFN, Sezione di Padova, via F Marzolo 8, I--35131, Padova, Italy.\\
        E-mail: \email{jae-hyeon.park@pd.infn.it}}
\author{W.~Porod\\
        Institut f\"ur Theoretische Physik und Astrophysik, Universit\"at W\"urzburg, D-97074 W\"urzburg, Germany.\\
        E-mail: \email{porod@physik.uni-wuerzburg.de}}
\author{O. Vives\\
        Departament de F\'{\i}sica Te\`orica and IFIC, Universitat de Val\`encia-CSIC, E-46100, Burjassot, Spain.\\
        E-mail: \email{oscar.vives@uv.es}}

\abstract{In order to satisfy current FCNC and CP violation bounds, SUSY flavour structures cannot be generic. An interesting solution to these SUSY Flavour and CP Problems lies on the use of an SU(3) family symmetry which spontaneously breaks CP symmetry. Typical observables of such a model are electric dipole moments and LFV processes. In addition, these models can give contributions to CP violation in neutral meson mixing. We show how the latter can be used to restrict the allowed SUSY parameter space, and induce correlations between LFV and EDM predictions.}

\FullConference{The 2009 Europhysics Conference on High Energy Physics,\\
		 July 16 - 22 2009\\
		 Krakow, Poland}

\begin{document}

\section{Introduction}

In most phenomenological introductions to the MSSM one learns that the model is haunted by the `SUSY Flavour and CP Problems.' Such problems are related to the SUSY contributions to flavour and CP violating processes, which turn out to be too large. However, when taking a closer look at both issues, the nature of each problem can be easily understood.

The \textit{SUSY flavour problem} arises when assuming that the unknown flavour-violating SUSY-breaking terms are generic, i.~e.~they are all of the same order.

The \textit{SUSY CP problem} arises when assuming that one can have large phases in flavour-independent parameters.

Nevertheless, the structure of the CKM matrix and the hierarchy of the fermion masses imply that the Yukawa matrices of the SM cannot be generic either. This strongly questions the assumption of generic flavoured SUSY-breaking terms, and suggests they should be structured. Furthermore, within the SM, the CKM phase is located within the flavour sector, which hints that maybe we should constrain all of the MSSM phases within the flavour sector too.

Our aim should then be to build a SUSY model capable of giving structure to all flavoured terms, while constraining all CP violation within the flavour sector. An attractive mechanism to achieve this aim is based on the breaking of a family symmetry with spontaneous CP violation. We shall concentrate on the $SU(3)$ model of~\cite{Ross:2004qn}.

\section{The RVV Model}

The RVV model is an effective SUSY model extended by an $SU(3)$ family symmetry, exact CP symmetry, and some additional $U(1)$ or discrete symmetries used to forbid unwanted terms. All fermion superfields are defined as triplets of $SU(3)$, such that Yukawa couplings are not allowed when the symmetry is unbroken. As additional fields, we have four flavons: $\theta_3$, $\overline\theta_3$, $\theta_{23}$ and $\overline\theta_{23}$. The (un)barred flavons are defined as (anti)triplets of $SU(3)$, so it is possible to build effective couplings between the fermion, flavon and Higgs superfields. The model also includes massive messenger superfields, which set the scale of the effective coupling. Once the effective couplings are defined, the flavons acquire vevs, and by this the Yukawa couplings are generated. The alignment of the flavon vevs and the ratio between them and the messenger masses determine then the hierarchical structure of the Yukawa matrices.

As a final detail, in order to differentiate the leptons from the quarks, the model also includes a Georgi-Jarlskog superfield $\Sigma$, which will acquire a vev in the $(B-L+2T^R_3)$ direction.

An effective superpotential capable of reproducing the observed quark mass hierarchy and mixing, with arbitrary order one coefficients, is:
\begin{align}
 W_{\rm Y} = H\psi _{i}\psi _{j}^{c} &\left[ \theta_{3}^{i}\theta_{3}^{j} + \theta _{23}^{i} \theta _{23}^{j}\Sigma 
+ \left(\epsilon ^{ikl} \overline{\theta}_{23,k} {\overline{\theta }_{3,l}}\theta_{23}^{j} 
+  \epsilon^{jkl}\overline{\theta}_{23,k} {\overline{\theta}_{3,l}}\theta _{23}^{i}\right) 
\left(\theta _{23} {\overline{\theta} _{3}}\right) + \right. \nonumber \\
 & \left.  \epsilon ^{ijl} \overline{ \theta}_{23,l} \left(
 \theta _{23} {\overline{\theta} _{3}}\right)^2 +  \epsilon ^{ijl} \overline{
 \theta}_{3,l} \left(\theta _{23} {\overline{\theta} _{3}}\right)\left(\theta
 _{23} {\overline{\theta} _{23}}\right) + \dots \right],
\end{align}
where for simplicity we have normalized $\theta_i=\theta_i/M_\psi$, and $M_\psi$ being the heavy messenger mass. The exact vacuum structure of the vevs can be found in~\cite{Ross:2004qn,Calibbi:2009ja}.

A feature of this model is the breaking of CP symmetry by the flavon vevs. The phases of the vevs cannot be all absorbed by a redefinition of the superfields, and thus remain as CP violating phases. Then the only possible complex parameters must be those coming from the effective couplings with the flavons, and by this all CP violation is constrained within the flavour sector.

The same mechanism for generating the Yukawas can be used in the MSSM to generate the structure of the SUSY-breaking terms. However, as the superpotential is a holomorphic function, while the SUSY-breaking terms are non-holomorphic, the textures can be different. The minimal structure that the SUSY-breaking terms can have is:
\begin{align}
\label{eq:softminimal}
(M^2_{\tilde f})_i^{j} = m_0^2 \bigg(\delta_i^j &
  +\left[\theta _{3,i}^{\dagger}\theta _3^j+\overline{\theta}_{3,i}{\overline{\theta}_{3}^\dagger}^j + 
\theta _{23,i}^{\dagger }\theta_{23}^j+\overline{\theta}_{23,i} 
{\overline{\theta}_{23}^\dagger}^j\right] \nonumber \\
& +(\epsilon_{ikl}\theta_3^k\theta_{23}^l)(\epsilon^{jmn}\theta_{3,m}^\dagger\theta_{23,n}^\dagger) 
+ (\epsilon_{ikl}{\overline{\theta }_3^\dagger}^k{\overline{\theta }_{23}^\dagger}^l)(\epsilon^{jmn}
\overline{\theta }_{3,m}\overline{\theta }_{23,n})+\quad\ldots\quad\bigg).
\end{align}
where we again have normalized the flavons. The detailed structures for the soft masses and trilinears in the SCKM basis, as well as non-minimal terms, can be found in~\cite{Calibbi:2009ja}.

It is important to notice that we have the same flavons and messengers as in the Yukawa sector. This means that the suppression factors and phases within the flavoured parameters of the MSSM shall be uniquely defined by the suppression factors and phases of the SM, the only difference lying in real, unknown $O(1)$ constants at $M_{GUT}$. These constants make it impossible for the model to predict exact values for any SUSY process, but determine unambiguously the associated order of magnitude.

\section{Flavoured Phenomenology}

With the determination of the order of magnitude and phase structure of each flavour violating term, it is then possible to make predictions for observables in the quark and lepton sector. In the following, it is important to take into account that all expectations can vary at most by a factor 2, or $0.5$, per mass-insertion, due to the unknown $O(1)$ constants at $M_{GUT}$.

\begin{figure}
\includegraphics[scale=.35]{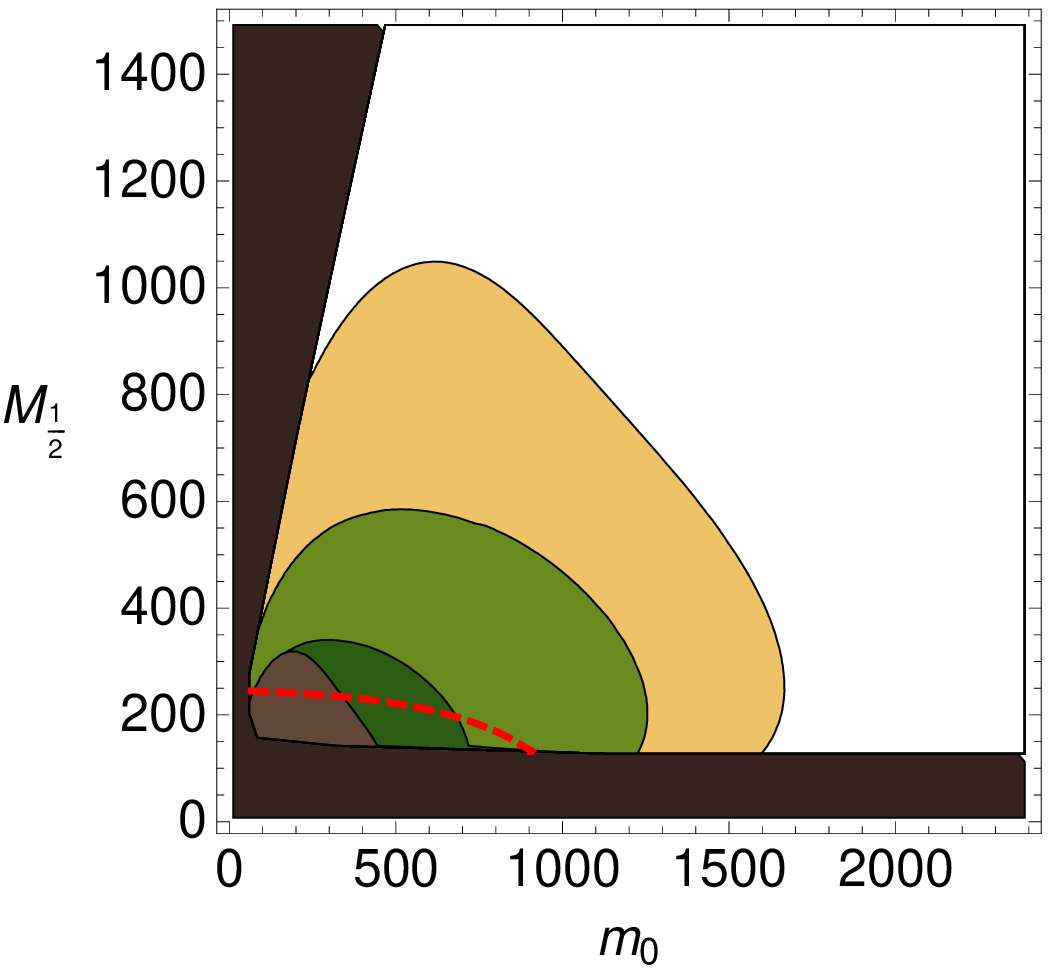} 
\includegraphics[scale=.35]{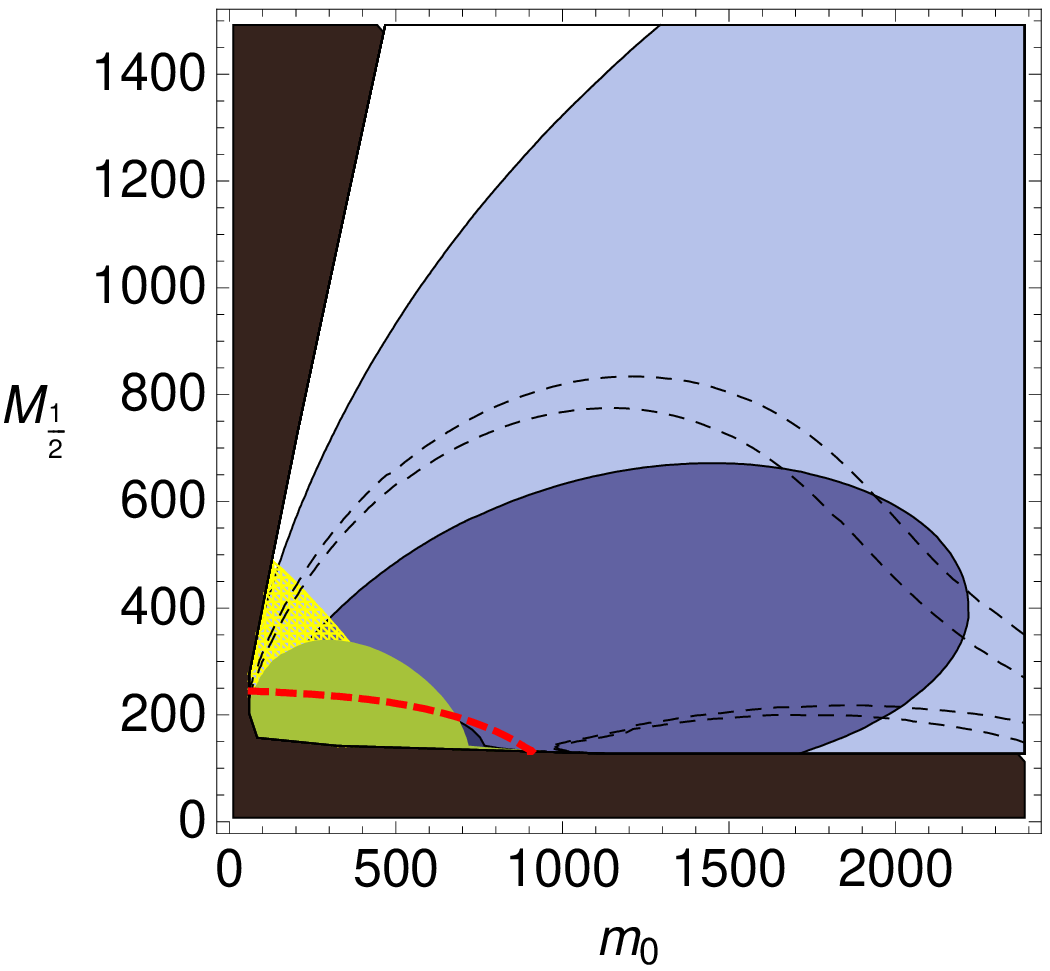}
\includegraphics[scale=.35]{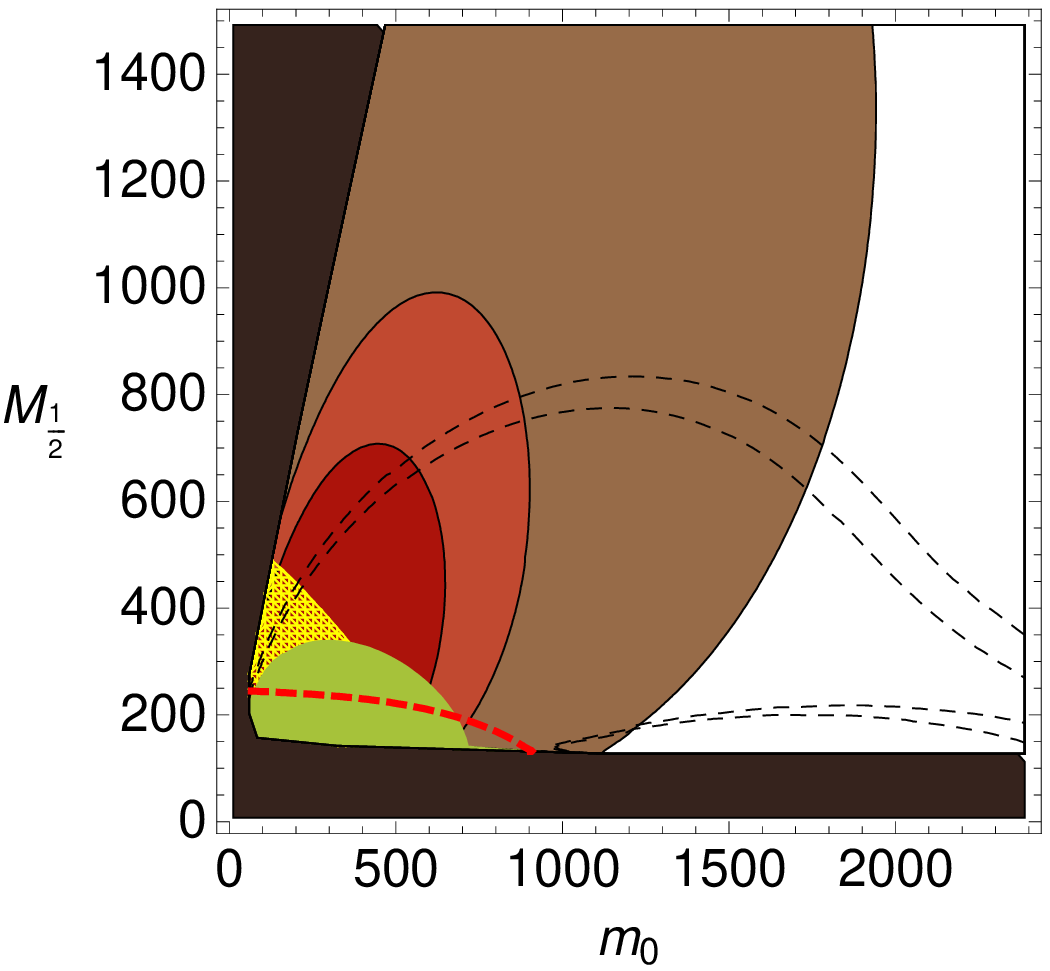} 
\includegraphics[scale=.35]{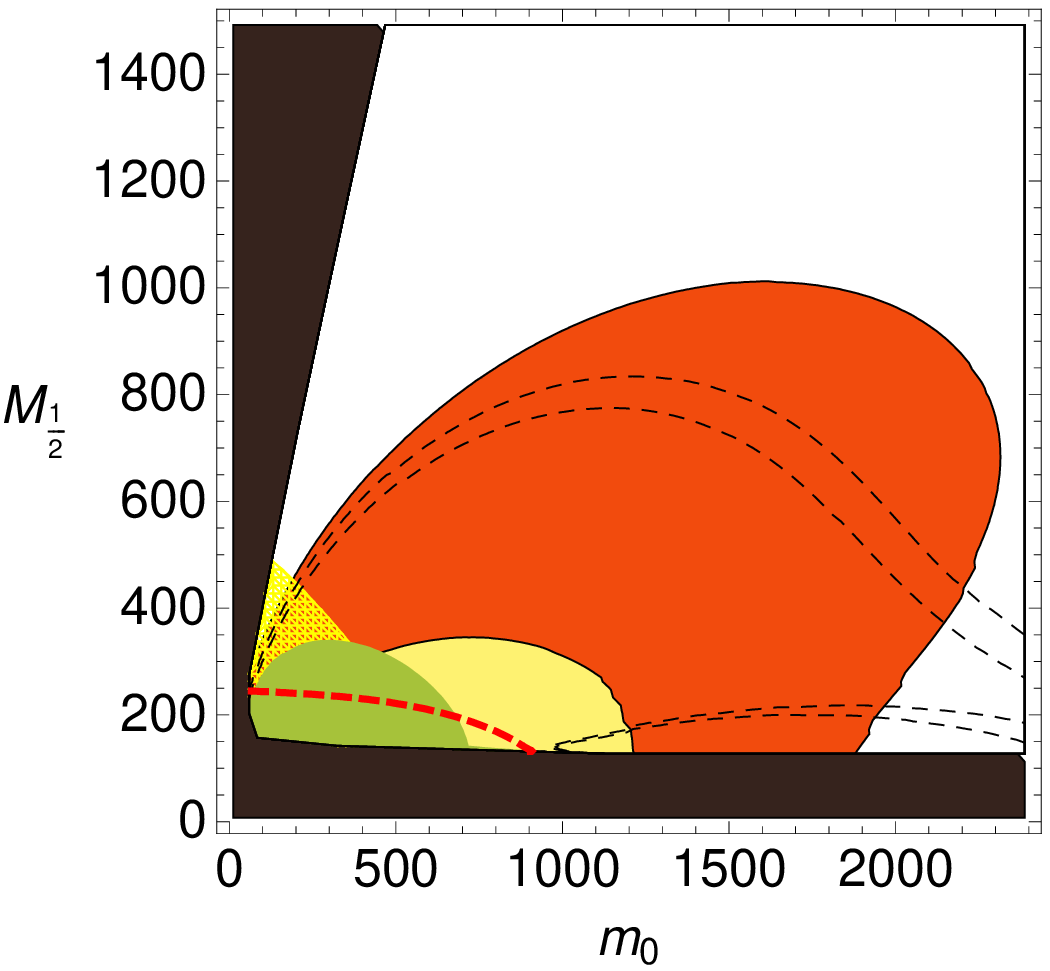}
\caption{\textit{(Far left)} Current and future LFV constraints. \textit{(Left)} Contours of $\phi_{B_s}$. \textit{(Right)} Future eEDM constraints. \textit{(Far right)} Future nEDM constraints. See the text for a description of each plot.}
\label{fig1}
\end{figure}

Figure~\ref{fig1} shows such predictions for LFV, meson mixing and EDMs, in the $m_0$-$M_{1/2}$ plane. We take $\tan\beta=10$ and $A_0=0$. In the plots, the dark brown region is excluded by having a charged LSP or by LEP, excepting the Higgs mass bound, which is shown in thick dashed, red lines.

The far left 
of Figure~\ref{fig1} shows our expectations for $\mu\to e\gamma$ ($\tau\to\mu\gamma$). The brown (green) regions correspond to the current sensitivity of the MEGA (BaBar+Belle) experiments. We can see that the current bounds constrain very slightly the parameter space. In contrast, the light brown (light green) regions show the future sensitivity of the MEG (Super Flavour Factory) experiments. In particular, it is noticeable that the region probed by MEG is roughly similar to the region probed by the LHC. This means that if SUSY is observed at the LHC, the observation of $\mu\to e\gamma$ is a likely signature of this model. In the following, we shall show the current LFV constraints in light green.

The left 
of Figure~\ref{fig1} shows contours for $\phi_{B_s}$, as defined in~\cite{Bona:2008jn}. The $(g-2)_\mu$ favoured region is also shown in hatched yellow. The light blue (dark blue) region shows $\phi_{B_s}$ larger than $10^{-5}$ ($10^{-4}$). It is unfortunate that the minimal version of this model cannot provide the required contribution to $\phi_{B_s}$ in order to solve the discrepancy reported in~\cite{Bona:2008jn}. This means that if the future data confirms the discrepancy, non-minimal versions of the model shall be required.

In addition to $\phi_{B_s}$, we show in this and all subsequent plots, the region of the parameter space capable of solving the $\epsilon_K$ tension reported by~\cite{Buras:2009pj}. It is shown as a strip between dashed black lines. It is important to emphasize that, since the $\epsilon_K$ tension is an $O(1)$ effect and not an order of magnitude effect, the location of the strip can vary much depending on the selected $O(1)$ constants. Nevertheless, it is encouraging to notice that the SUSY contribution from the model has the required order of magnitude to solve the tension.

The right and far right of Figure~\ref{fig1} shows the order of magnitude of the electron and neutron EDMs, respectively. The current bounds do not constrain the parameter space. We show expectations larger than $10^{-29}$, $5\cdot10^{-30}$, $10^{-30}$ in dark red, red and brown for the eEDM, and larger than $10^{-28}$, $10^{-29}$ in light yellow and orange for the nEDM. Thus, if we require solving the $(g-2)_\mu$ and $\epsilon_K$ problems, we expect the eEDM to be observed in the next experiments, but not the nEDM.

\section{Conclusions}

We have presented an $SU(3)$ model~\cite{Ross:2004qn} capable of generating the structure of the CKM matrix, as well as the observed hierarchy in the fermion masses. When applied within the MSSM, the model provides textures to the SUSY-breaking terms, solving the SUSY flavour and CP problems.

Although the minimal version of the model is not capable of generating a large enough value for $\phi_{B_s}$~\cite{Bona:2008jn}, it has the ability to accommodate the $\epsilon_K$ tension~\cite{Buras:2009pj}. In addition, it is expected to observe LFV processes and an eEDM in the next series of experiments. The nEDM is somewhat small to be observed in the near future.

It is of interest to note that the definition of an $\epsilon_K$ strip can impose correlations between all of the aforementioned observables~\cite{Calibbi:2009ja}. This is important in order to distinguish between different flavour symmetry models.


\begin{thebibliography}{99}

\bibitem{Ross:2004qn}
G.~G.~Ross, L.~Velasco-Sevilla and O.~Vives,
Nucl.\ Phys.\ B {\bf 692}, 50 (2004)
[arXiv:hep-ph/0401064]

\bibitem{Calibbi:2009ja}
  L.~Calibbi, J.~Jones-Perez, A.~Masiero, J.~h.~Park, W.~Porod and O.~Vives,
  arXiv:0907.4069 [hep-ph].

\bibitem{Bona:2008jn}
  M.~Bona {\it et al.}  [UTfit Collaboration],
  arXiv:0803.0659 [hep-ph].

\bibitem{Buras:2009pj}
  A.~J.~Buras and D.~Guadagnoli,
  arXiv:0901.2056 [hep-ph].

\end{thebibliography}
\end{document}